\documentclass[
aps,
twocolumn,
superscriptaddress,
prl,
amsmath,amssymb
]{revtex4-1}

\usepackage{graphicx}
\usepackage{dcolumn}
\usepackage{bm}
\usepackage{color}
\usepackage{hyperref}

\begin{document}

\title{Torsional optomechanics of a levitated nonspherical nanoparticle}

\author{Thai M. Hoang}
 \affiliation{Department of Physics and Astronomy, Purdue University, West Lafayette, IN 47907, USA}

 \author{Yue Ma}
 \affiliation{Center for Quantum Information, Institute of Interdisciplinary Information Sciences, Tsinghua University, Beijing, 100084, China}
\affiliation{Department of Physics, Tsinghua University, Beijing 100084, China}

\author{Jonghoon Ahn}
 \affiliation{School of Electrical and Computer Engineering, Purdue University, West Lafayette, IN 47907, USA}

\author{Jaehoon Bang}
 \affiliation{School of Electrical and Computer Engineering, Purdue University, West Lafayette, IN 47907, USA}

\author{F. Robicheaux}
 \affiliation{Department of Physics and Astronomy, Purdue University, West Lafayette, IN 47907, USA}
 \affiliation{Purdue Quantum Center, Purdue University, West Lafayette, IN 47907, USA}

\author{Zhang-Qi Yin}
\email{yinzhangqi@mail.tsinghua.edu.cn}
 \affiliation{Center for Quantum Information, Institute of Interdisciplinary Information Sciences, Tsinghua University, Beijing, 100084, China}

\author{Tongcang Li}
 \email{tcli@purdue.edu}
 \affiliation{Department of Physics and Astronomy, Purdue University, West Lafayette, IN 47907, USA}
 \affiliation{School of Electrical and Computer Engineering, Purdue University, West Lafayette, IN 47907, USA}
 \affiliation{Purdue Quantum Center, Purdue University, West Lafayette, IN 47907, USA}
 \affiliation{Birck Nanotechnology Center, Purdue University, West Lafayette, IN 47907, USA}

\date{\today}

\begin{abstract}
An optically levitated nanoparticle in vacuum is a paradigm optomechanical system for sensing and studying macroscopic quantum mechanics.
While its center-of-mass motion  has been investigated intensively, its torsional vibration has only been studied theoretically in limited cases.
Here we report the first experimental observation of the  torsional vibration of an optically levitated nonspherical nanoparticle in vacuum. We achieve this by utilizing the coupling between the spin angular momentum of photons and the torsional vibration of a nonspherical nanoparticle whose polarizability
is a tensor. The torsional vibration frequency can be one order of magnitude higher
than its center-of-mass motion frequency, which is promising for ground state cooling. We propose a simple yet novel scheme to achieve ground state cooling of its torsional vibration with a linearly-polarized Gaussian cavity mode. A levitated nonspherical nanoparticle in vacuum will also be an ultrasensitive nanoscale torsion balance with a torque detection sensitivity on the order of $10^{-29} ~\mathrm{N}\cdot \mathrm{m}/\sqrt{\mathrm{ Hz}}$ under realistic conditions.
\end{abstract}
\maketitle

An optically levitated dielectric particle in vacuum \cite{ashkin1976optical, Li2011, yin2013review} is an ultrasensitive detector for force sensing \cite{ranjit2015attonewton,ranjit2016zeptonewton}, millicharge searching \cite{moore2014search} and other applications \cite{Arvanitaki2013, rider2016search}.
It will provide a great platform to test fundamental theories such as objective collapse models \cite{RomeroIsart2011, romero2011quantum} and quantum gravity \cite{albrecht2014testing} when its mechanical motion can be cooled to the quantum regime \cite{Romero10, Chang2010Cavity}.
Recently, feedback cooling of the center-of-mass (COM) motion of a levitated nanosphere to about 450 $\mu$K (about 63 phonons at 150 kHz) \cite{jain2016direct}, and cavity cooling of the COM motion of a nanosphere  to a few mK \cite{fonseca2015nonlinear} were demonstrated.
The vibration mode would have already been  in ground state at 450~$\mu$K \cite{jain2016direct} if its frequency is above 10~MHz.
Increasing the vibration frequency of the nanoparticle can be a key to achieve ground state cooling.
However, this can not be achieved by simply increasing the intensity of the trapping laser, which induces heating and subsequently causes the loss of the  nanoparticle \cite{millen2014nanoscale,ranjit2015attonewton}.
Besides COM motion, a pioneering work has proposed to use multiple Laguerre-Gaussian (LG) cavity modes to achieve angular trapping of a dielectric rod and cool its torsional vibration (TOR)  to the ground state \cite{Romero10}.
This  was later generalized to micro-windmills \cite{shi2013coupling}, which have better overlap with LG cavity modes. 
These intriguing proposals of torsional optomechanics, however, have not been realized experimentally yet.

In this work, we report the first experimental observation of the  torsional vibration of an optically levitated nonspherical nanoparticle in vacuum, and show that the torsional frequency can be one order of magnitude higher than the COM frequency at the same laser intensity.
We explain our observation using a  model of an ellipsoidal nanoparticle levitated by a linearly-polarized Gaussian beam.
For an ellipsoid much smaller than the wavelength of the trapping laser, its polarizability is a tensor due to its  geometry \cite{Bohren2007particles}.
In a linearly polarized Gaussian beam, the long axis of an ellipsoid  tends to align with the polarization direction of the trapping laser to minimize the potential energy.
When its long axis deviates from the polarization direction of the trapping laser, the ellipsoid will experience a torque pushing its long axis back to the equilibrium orientation.
As a result, the ellipsoid will experience angular trapping and exhibit torsional vibration. Both the frequency and the quality factor of the torsional vibration can  be one order of magnitude higher than those of the COM motion.
Inspired by the experimental observation, we propose a simple yet novel scheme to achieve TOR ground state
cooling with a cavity driven by a linearly-polarized Gaussian beam. While we use nanodiamonds as examples in this paper, our proposals will also work for other transparent nonspherical nanoparticles \cite{kuhn2015cavity,lechner2013cavity}.

Besides being a platform for investigating fundamental physics, a levitated nonspherical nanoparticle in vacuum will also be a nanoscale torsion balance \cite{kim2013nanoscale,wu2014dissipative}. It can have a torque detection sensitivity on the order of $10^{-29}~\mathrm{N}\cdot \mathrm{m}/\sqrt{\mathrm{ Hz}}$ under realistic conditions, which will open up many new applications. Angular trapping and torsional vibration are also important for  spin-optomehcanics of levitated nanodiamonds with nitrogen-vacancy (NV) centers \cite{yin2013large,Scala2013,yin2015,Neukirch2015,hoang2015observation}, for which the orientations of NV centers are important.

\begin{figure}[btp]
	\includegraphics[scale=0.9]{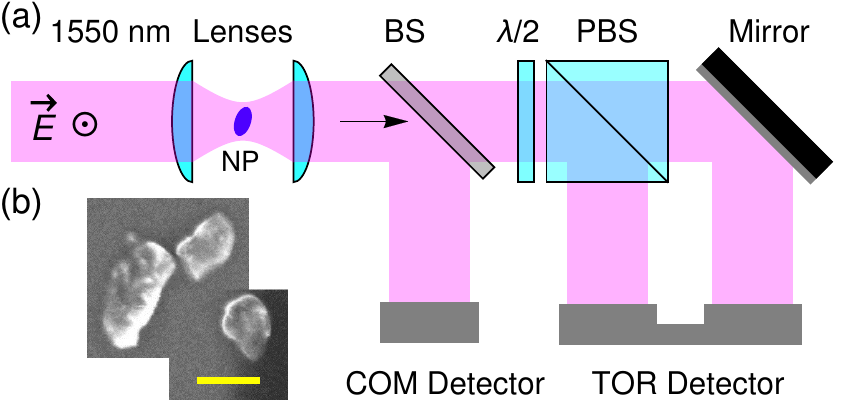}
\includegraphics[scale=0.9]{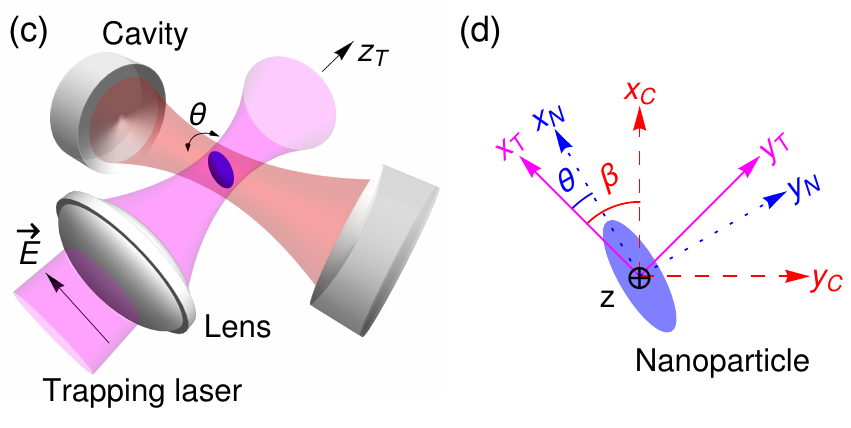}	
	\caption{(Color online) (a) Experimental diagram for detecting torsional (TOR) vibration and center-of-mass (COM) motion of a levitated nonspherical nanoparticle (NP). A nanodiamond (represented by an ellipsoid) is levitated by a tightly focused linearly-polarized 1550~nm laser beam.  The nanoparticle's motion is monitored by the   exiting trapping laser. The exiting beam is split by a beam splitter (BS) to a COM detector and a TOR detector. A $\lambda/2$ waveplate balances the power of the beams after the polarizing beam splitter (PBS) for the TOR detector. (b) An SEM image of irregular nanodiamonds. The scale bar  is 100~nm. (c) A proposed scheme to cool the torsional vibration of a levitated ellipsoidal nanoparticle with an optical cavity driven by a linearly polarized Gaussian beam. (d) The relation between  Cartesian coordinate systems of the nanoparticle ($x_N$, $y_N$, $z_N$), the trapping laser ($x_T$, $y_T$, $z_T$), and the cavity mode ($x_C$, $y_C$, $z_C$).
$x_N$ axis aligns with the longest axis of the nanoparticle. $x_T$ and $x_C$ axes align with the polarization directions of the trapping laser and the cavity mode, respectively. The angle between $x_N$ and $x_T$ is $\theta$, and the angle between $x_T$ and $x_C$ is $\beta$.
$y_C$ axis is the optical axis of the cavity.  $z_T$ axis is the propagation direction of the trapping laser. As we only consider one torsional mode, we assume $z_C$ and $z_N$ are parallel to $z_T$ for simplicity.
}	
\label{fig:scheme1}
\end{figure}

\begin{figure}
	\begin{minipage}{3.4in}
	\includegraphics[scale=0.95]{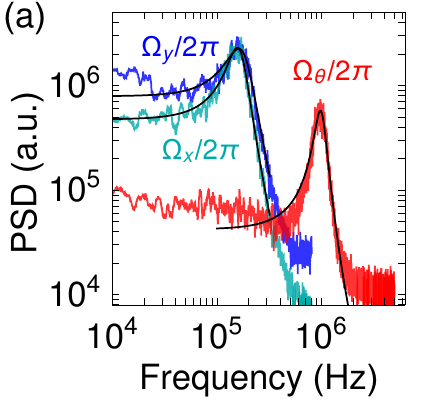}
	\includegraphics[scale=0.95]{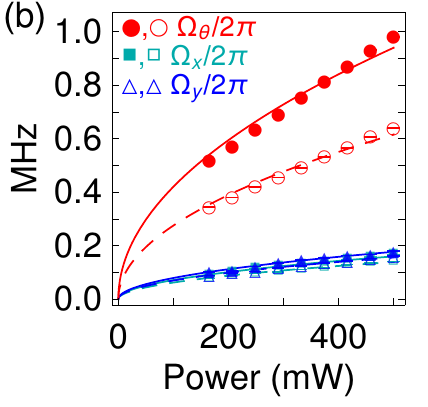}	
\\
	\includegraphics[scale=0.95]{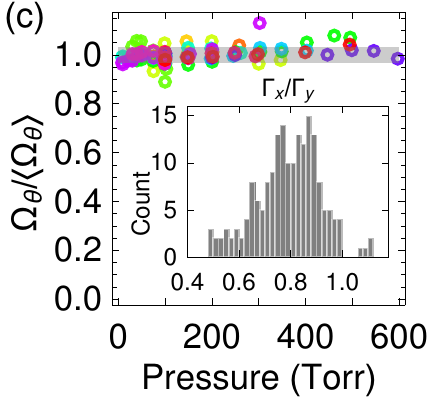}
	\includegraphics[scale=0.95]{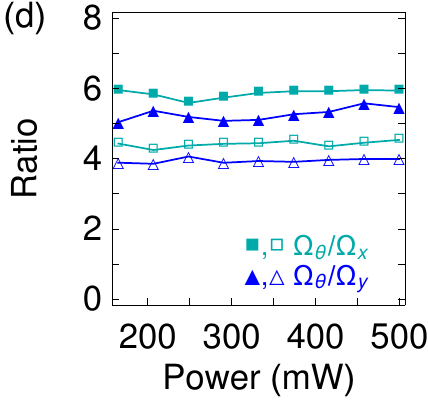}	\\
	\end{minipage}		
	\caption{ (Color online) (a) Measured power spectrum densities (PSD) of the COM motion (labeled with $\Omega_{x,y}/2\pi$) and TOR vibration  (labeled with $\Omega_\theta/2\pi$) of an optically levitated nanodiamond at 100 Torr. When the   power of the trapping laser is about 500~mW,  $\Omega_x/2\pi = 0.16~\mathrm{MHz}$, $\Omega_y/2\pi = 0.18~\mathrm{MHz}$ and  $\Omega_{\theta}/2\pi = 1.0~\mathrm{MHz}$. The damping coefficient is anisotropic with a ratio  $\Gamma_{x}/\Gamma_{y}=0.69$ for this nanodiamond.  (b) Measured TOR and COM frequencies  of two different nanodiamonds at 100 Torr as a function of the trapping power. Frequency data are fitted to a function $A\sqrt{P}$ with $P$ being the trapping power and $A$ being a fitting parameter. Solid and open symbols represent different nanodiamonds.  (c) $\Omega_{\theta}/ \langle \Omega_\theta \rangle$, of 13 nanodiamonds at different pressures. $\langle \Omega_\theta \rangle$ is the average over different pressures of a nanodiamond's torsional frequency. The shaded region is the standard deviation of all $\Omega_\theta/ \langle \Omega_\theta\rangle$ data. The inset plot is the histogram of the measured ratios of  damping coefficients in  $x-$ and $y-$directions ($\Gamma_x/\Gamma_y$) for these nanodiamonds  at many different trapping powers and pressures. $\Gamma_x/\Gamma_y<1$ indicates that the long axes of the nanodiamonds align with the polarization direction ($x$-axis) of the laser beam.   (d) The ratio $\Omega_\theta/\Omega_{x,y}$ of two  levitated nanodiamonds (shown in (b)) at different laser powers. a.u. denotes arbitrary unit. }	
	\label{Fig:exptorsional}
\end{figure}

\textit{Observation of the torsional vibration of levitated nanoparticles.}
In the experiment, nanodiamonds are levitated using an optical tweezer formed by a linearly-polarized 1550~nm laser beam (Fig. \ref{fig:scheme1}{a}).
The laser beam is tightly focused with a NA=0.85 objective lens \cite{hoang2015observation}. The nanodiamonds  have broad distributions around their manufacture size 100~nm. Some nanodiamonds have large aspect ratios as shown in Fig. \ref{fig:scheme1}{b}.
The torsional vibration of the nanodiamond will change the polarization of the laser beam, which can be detected with a polarizing beam splitter (PBS) and a balanced detector (Fig. \ref{fig:scheme1}{a}).
Similar detection schemes have been used to detect the  rotation of  birefringent particles driven by  circularly-polarized lasers \cite{Nieminen2001optical,Porta2004Optical,Arita2013laser}. The COM motion of the nanodiamond changes the direction of the laser beam and can be detected with a balanced detector  \cite{Li2010,li2012fundamental}.

A sample of the power spectrum density (PSD) of the COM motion and the TOR motion of a levitated nanodiamond is shown in Fig. \ref{Fig:exptorsional}{a}. Over 1/3 of our trapped nanodiamonds exhibit torsional signals.
For this nanodiamond, the TOR frequency ($\Omega_\theta/2\pi=1.0$~MHz) is about 6 times higher than the transverse COM frequency ($\Omega_x/2\pi=0.16$~MHz and $\Omega_y/2\pi=0.18$~MHz), which is promising for ground state cooling. For comparison, a factor of 6 increase in the COM frequency would require the laser power to be increased by a factor of 36, which can induce significant heating of the nanoparticle.

We investigated the  motions of many different nanodiamonds as a function of trapping powers (Fig. \ref{Fig:exptorsional}b,d) and air pressures (Fig. \ref{Fig:exptorsional}c).
For each nanodiamond exhibiting torsional vibration, the TOR frequency  is proportional to the square root of the trapping power, $\Omega_\theta \propto \sqrt{P}$, as shown in Fig. \ref{Fig:exptorsional}{b}. This observation agrees with the prediction of the ellipsoidal model discussed below.
Since the COM frequency is also proportional to the square root of the trapping power, $\Omega_y \propto \sqrt{P}$,   the ratio  $\Omega_\theta/\Omega_{x,y}$ is independent of the trapping power, as shown in 
Fig. \ref{Fig:exptorsional}{d}. 
The TOR frequency is independent of the air pressure as shown in Fig. \ref{Fig:exptorsional}{c} when the  pressure is reduced from atmospheric pressure to a few Torr.
The summary of 13~nanodiamonds yields $\Omega_\theta/\langle \Omega_\theta\rangle = 1 \pm 0.03 $ across the whole pressure range. Here $\langle \Omega_\theta\rangle$ is the averaged torsional frequency over different pressures for each particle.
For comparison, the rotational frequency of a birefringent microsphere driven by a circularly polarized laser is linearly proportional to the laser power, and increases when the air pressure decreases \cite{Arita2013laser}, which are very different from our results of the torsional vibration.  

 Further experimental evidence that supports the interpretation of torsional vibration instead of free rotation is from the measured damping factors which are anisotropic. The measured ratios, $\Gamma_x/\Gamma_y$, for all the nanoparticles in Fig. \ref{Fig:exptorsional}{c} at different pressures and trapping powers yield a mean value of $0.8$. A nonspherical  nanoparticle rotating in the $xy$-plane should yield $\Gamma_x/\Gamma_y=1$ on average \cite{Gieseler2012Subkelvin}. $\Gamma_{x}/\Gamma_y<1$ means  the long axes of the nanoparticles align with the polarization direction ($x$-axis) of the trapping laser \cite{li2012effect}.

\textit{An ellipsoidal model.}
As a minimal model to describe the torsional vibration of an irregular nanodiamond (Fig. \ref{fig:scheme1}{b}), we consider an ellipsoid with semiaxes $r_x>r_y=r_z$ in a linearly-polarized optical tweezer.
When the size of the ellipsoid is much smaller than the wavelength  of the  laser, we can use the Rayleigh approximation. The induced dipole will be $\textbf{p}= \alpha_x E_x \hat{x}_N + \alpha_y E_y \hat{y}_N+\alpha_z E_z \hat{z}_N$, where the instantaneous electric field of the laser beam, $\textbf{E}$, is  decomposed into components along the principle axes of the ellipsoid \cite{Bohren2007particles}. Here $\alpha_x$, $\alpha_y$, $\alpha_z$ are polarizabilities along the principle axes.
The force and torque on the ellipsoid are $\textbf{F} = \langle \nabla( \textbf{p} \cdot \textbf{E}) \rangle /2$ and $\textbf{M} = \langle \textbf{p} \times \textbf{E} \rangle $, respectively \cite{Bohren2007particles,Trojek2012Optical}. To capture the essential properties of the system,  we only consider the COM motion of the ellipsoid along the $y_T$ axis and the torsional vibration of the ellipsoid around the $z_T$ axis (Fig. \ref{fig:scheme1}d).
The potential energy of the ellipsoid in the optical tweezer is:
\begin{equation}\label{eq:potential}
U (y,\theta)= -\frac{V}{2 c} [\chi_x-(\chi_x-\chi_y)\sin^2\theta] I_{L}(y),
\end{equation}
where $V=4\pi r_x r_y r_z/3$ is the volume of the ellipsoid, $c$ is the speed of light, $\chi_x=\alpha_x/(\epsilon_0 V)$ and $\chi_y=\alpha_y/(\epsilon_0 V)$ are the effective  susceptibility of the ellipsoid, $\epsilon_0$ is the vacuum permittivity, $\theta$ is the angle between the longest axis of the ellipsoid and the electric field of the laser beam,  and $I_{L}(y)$ is the laser intensity at the location of the ellipsoid. As an example, $\chi_x=2.05$, $\chi_y=1.74$ for an ellipsoidal nanodiamond with $r_y/r_x=0.8$.

\begin{figure}[btp]
\setlength{\unitlength}{1cm}
\includegraphics[totalheight=6.6cm]{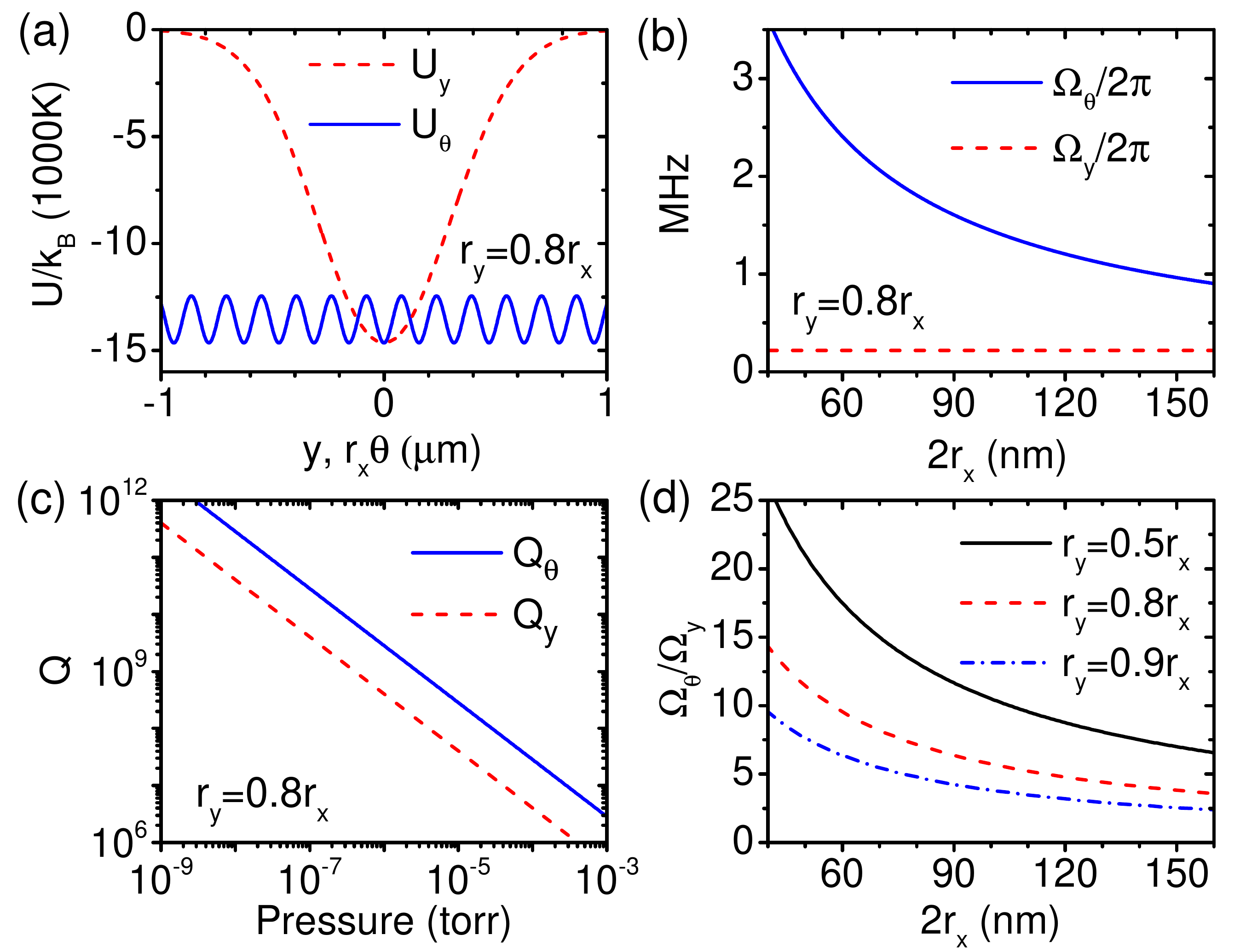}
\caption{(Color online) (a) Trapping potentials in the transverse ($U_{y}$) and angular ($U_{\theta}$) directions. (b) Frequencies of the TOR and COM vibrations as a function of the size of the ellipsoid when its aspect ratio is $r_y/r_x=0.8$. (c) Quality factors of the TOR vibration and COM motion of a levitated nanoparticle as a function of the pressure. (d) Enhancement ratio $\Omega_{\theta}/\Omega_y$ as a function of the size of the ellipsoid with different aspect ratios.   In the calculations, we assume the waist of the Gaussian optical tweezer to be 600 nm, the laser wavelength to be 1550 nm, and the laser power to be 100 mW.  To calculate results shown in subfigures (a) and  (c), we assume the semiaxes of the ellipsoid to be  $r_x=50$~nm,  $r_y=r_z=40$~nm. The corresponding vibration  frequencies are  $\Omega_\theta/2{\pi}=1.26$~MHz and $\Omega_y/2{\pi}=220$~kHz. }
\label{fig:ellipsoidtheory}
\end{figure}

The trapping potential $U_y$ as a function of the  transverse position $y$ of an ellipsoid with  orientation  $\theta=0$, and the potential $U_{\theta}$ as a function of $r_x\theta$ when $y=0$ are shown in Fig. \ref{fig:ellipsoidtheory}{a}. The semiaxes of the diamond ellipsoid are   $r_x=50$~nm and  $r_y=r_z=40$~nm.
To plot  $U_\theta$ and $U_y$ together, we use $r_x\theta$, the distance the apex point moves, as the horizontal axis for $U_\theta$.
While the COM potential energy $U_y$ has a Gaussian shape reflecting the spatial profile of the Gaussian laser beam, the torsional potential energy $U_\theta$ is a sinusoidal function of the angle $\theta$ as illustrated in Eqn. \ref{eq:potential}.
$U_\theta$ has several periods within the range of $|r_x\theta| < W_t$. Here $W_t$ is the waist of the trapping laser. Thus we expect the torsional vibration to have a higher frequency than that of the COM motion.
The frequencies of the  TOR vibration and COM motion of the ellipsoid calculated from the potential energy (Eqn. \ref{eq:potential}) are
{\small
\begin{eqnarray}
		\Omega_{y} = \sqrt{\frac{4\chi_x P}{c\pi \rho W^4_t}}
		, ~~\Omega_{\theta} = \sqrt{\frac{10(\chi_x - \chi_y)P}{c\pi\rho W^2_t (r_x^2 + r_y^2)}}.
	\label{Eqn:COMTORfreq}
\end{eqnarray}
}Here $\rho$ is the particle mass density, $P$ is the power of the Gaussian trapping laser.
The  ratio $\Omega_{\theta}/\Omega_y$ scales as $W_t/r_x$ with the semiaxis of the nanoparticle $r_x$ typically $10-20$ times smaller than the laser waist $W_t$.

While the  frequency of the COM motion is independent of the size of the ellipsoid, the  frequency of the torsional vibration increases when the size of the ellipsoid decreases (Fig. \ref{fig:ellipsoidtheory}{b}).
 Both experimental (Fig. \ref{Fig:exptorsional}) and theoretical results (Fig. \ref{fig:ellipsoidtheory}{b,d}) demonstrate that the TOR frequency can be one order of magnitude higher than the COM frequency at the same  laser intensity. The ratio $\Omega_\theta/\Omega_y$ can be  increased by decreasing the  ratio  $r_y/r_x$ or  the particle size. Because of a higher frequency, the quality factor of the TOR vibration ($Q_{\theta}$) \cite{halbritter1974} can also be one order of magnitude higher than that of the COM motion ($Q_{y}$) \cite{li2012effect}, which is another advantage for torsional ground state cooling (Fig.\ref{fig:ellipsoidtheory}{c}).
In  the calculation, we assume the collisions between  air molecules and the ellipsoidal nanoparticle are mainly inelastic, with a momentum accommodation coefficient  of 0.9 \cite{halbritter1974,li2012effect}.

\textit{Torque sensing.} An optically levitated ellipsoid in vacuum will be an ultrasenstive nanoscale torsion balance \cite{cavendish1798experiments} using the  laser as a ``string'' to provide the restoring torque. Torsion balances have played historic roles in the development of modern physics. They were used in the Coulomb's experiment that discovered the law of electrostatic force,  the Cavendish experiment that measured the gravitational constant \cite{cavendish1798experiments}, and many other important experiments\cite{eotvos1890,beth1936mechanical,adelberger2009torsion}. The minimum  torque that can be detected with a torsion balance is $M_{min}=\sqrt{4 k_B T I \Omega_{\theta}/(Q_{\theta}\Delta t)}$ \cite{haiberger2007highly}. Here  $T$ is the environmental temperature, $I$ is the moment of inertia, and $\Delta t$ is the measurement time. For a levitated $r_x=50$~nm,  $r_y=r_z=40$~nm ellipsoid  at $10^{-8}$~Torr  with a torsional frequency of $\Omega_\theta/2\pi=1.26$~MHz, the torque sensitivity is about $2 \times 10^{-29} ~\mathrm{N}\cdot \mathrm{m}/\sqrt{\mathrm{ Hz}}$ at 300~K. This is  several orders more sensitive  than tethered nanoscale torque sensors, which typically have sensitivities on the order of $10^{-21} ~\mathrm{N}\cdot \mathrm{m}/\sqrt{\mathrm{ Hz}}$  \cite{kim2013nanoscale,wu2014dissipative}. This system can be used to measure the torque on a single electron spin\cite{rugar2004single} or even a single nuclear spin. A proton in a 0.1~Tesla magnetic field would experience a torque  on the order of $10^{-27} ~\mathrm{N}\cdot \mathrm{m}$.

\begin{figure}[tb]
\includegraphics[scale=0.3]{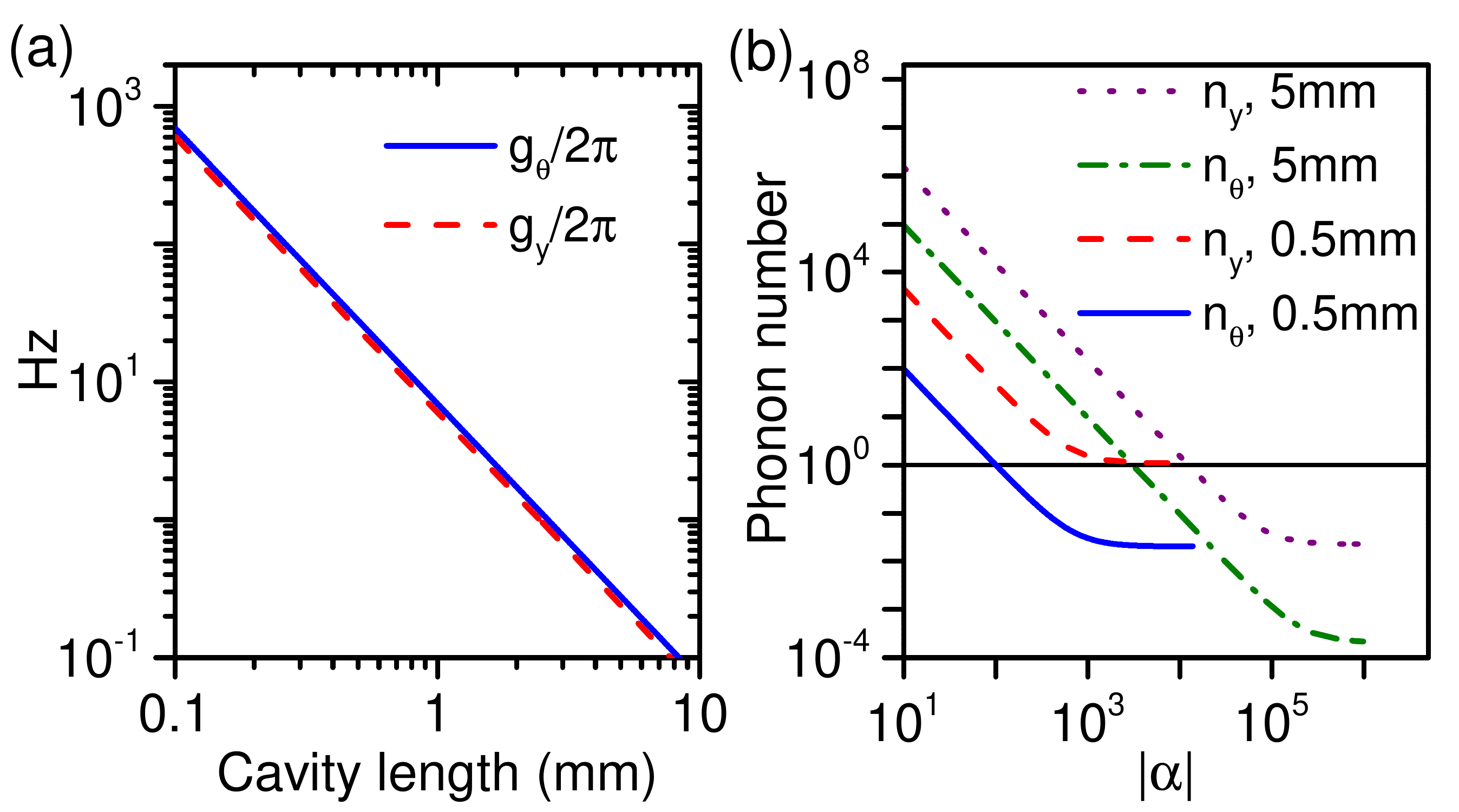}
\caption{(color online) (a) The single phonon-single photon coupling strengths of the COM motion ($g_y$) and torsional vibration $g_{\theta}$ as a function of cavity length.  For the calculation, we assume the size of the diamond ellipsoid to be $r_x=50$~nm, and  $r_y=r_z=25$~nm.  The trapping laser parameters are the same as in Fig. \ref{fig:ellipsoidtheory}. The calculated vibration frequencies are $\Omega_{\theta}/2{\pi}=2.6$ MHz and   $\Omega_y/2{\pi}=248$ kHz.  (b)  The final phonon numbers of the COM mode ($n_y$) and the torsional mode ($n_{\theta}$) as a function of the driving field.  The resonant wavelength of the cavity is 1540~nm, which is different from that of the trapping laser. The background pressure is assumed to be $10^{-8}$ Torr. The cavity finesse is  $\mathcal{F}=10^5$.}
\label{fig:cavitycooling}
\end{figure}

\textit{Torsional ground state cooling.}
Inspired by the experimental observation and the ellipsoidal model, we propose to use a linearly-polarized Gaussian beam to drive a cavity to cool the torsional vibration to the ground state \cite{Wilson2007Theory, Chang2010Cavity, Romero10}. We assume the nonspherical nanoparticle is levitated using another linearly-polarized Gassian beam (Fig. \ref{fig:scheme1}{c, d}).
Because the TOR vibration and COM motion can have very different frequencies, we can neglect their coupling.  For simplicity, we only consider the torsional mode $\Omega_{\theta}$ around the $z_T$ axis.  The linear Hamiltonian of the system in the frame rotating at the cooling laser frequency can be approximated as  \cite{Wilson2007Theory, Romero10}
\begin{eqnarray}
	\hat{H} = -\hbar \Delta_L \hat{a}^\dag \hat{a} + \hbar \Omega_{\theta} \hat{b}^{\dag}\hat{b} + \hbar|\alpha| g_{\theta}(\hat{b}^{\dag} + \hat{b})(\hat{a}^\dag + \hat{a}).
\end{eqnarray}
Here $2\pi \hbar$ is the Planck constant.  $\Delta_L=\omega_L-\omega_C+ 2 g^2_{\theta} |\alpha|^2 /\Omega_{\theta}$ is the effective detuning. $\omega_L$ is the laser frequency,   $\omega_C$ is the cavity resonant frequency. $\hat{a}^\dag (\hat{a})$ and $\hat{b}^\dag (\hat{b})$ are the creation (annihilation) operators for the cavity field and the mechanical motion, respectively. $|\alpha|=\sqrt{n_{p}}$ is the steady  amplitude of the cavity mode. $n_p$ is the number of photons in the cavity.
$g_{\theta}$ is the coupling strength between a single torsional vibration phonon  and a single cavity photon \cite{Wilson2007Theory}.
The coupling strength $g_{\theta}$ will be maximized when the center of the nanoparticle is at the antinode of the cavity mode, and  the angle between the polarization directions of the trapping laser and the cavity beam is  $\beta=45^{\circ}$ (Fig. \ref{fig:scheme1}{d}).
We obtain the maximum coupling constant as \cite{Wilson2007Theory, Romero10}
\begin{eqnarray}
	g_\theta &=& \sqrt{\frac{10\hbar \pi r_x r_y^2}{3\rho(r_x^2 + r_y^2)\Omega_\theta}}(\chi_x -\chi_y) \frac{64\pi c}{\lambda^2_C L^2}.
\end{eqnarray}
Here $\lambda_C$ is the wavelength of cavity mode, and  $L$ is the length of the cavity. The waist of the cavity mode is $W_C=\sqrt{\lambda_C L/2\pi}$ for a confocal cavity. 

The single phonon-single photon coupling strength $g_{\theta}$  as a function of the cavity length is shown in Fig. \ref{fig:cavitycooling}{a}.
For comparison, we also plot the maximum coupling strength  between the COM motion and the cavity mode $g_y= \sqrt{\frac{2\hbar \pi r_x r_y^2}{3\rho\Omega_y}}\chi_x \frac{16\pi^2 c}{\lambda^3_C L^2}$, which happens when $\beta=0^{\circ}$.
For a $r_x=50$~nm, and  $r_y=r_z=25$~nm nanodiamond, $g_{\theta}$ and $g_y$ have similar magnitudes and both increase when the length of the cavity decreases. Using a similar procedure in Refs. \cite{Wilson2007Theory,Wilson2007Cavity}, we can calculate the steady state phonon number of the levitated nanoparticle.
We assume the finesse of the cavity is $\mathcal{F}=10^5$ and the detuning is $\Delta_L=-\sqrt{\frac{\kappa^2}{4}+\Omega_{\theta}^2}$ with $\kappa$ being the decay rate of the cavity.
 As shown in Fig. \ref{fig:cavitycooling}{b},  only the torsional mode can be cooled to the ground state with a $L=0.5$~mm cavity. For a $L=5$~mm cavity, both the torsional mode and the COM mode can be cooled to the ground states. The torsional vibration mode will have smaller final phonon numbers because of its higher vibration frequency.
Thus the torsional mode can be cooled to the ground state with a broader range of cavity length and driving field strength than those of the COM mode.
 Comparing to former proposals with high-order LG cavity modes \cite{Romero10,shi2013coupling,shi2015optomechanics}, our proposal only requires linearly-polarized Gaussian beams. The observed motion is similar to the torsional vibrations (``pendular states'') of molecules \cite{friedrich1991spatial,stapelfeildt2003rmp} and spins \cite{Jing2011prl} in an external field, and 
 can be used to study torsional decoherence \cite{zhong2016}.

This material is based upon work supported by the National Science Foundation under Grant No. 1555035-PHY and Grant No. 1404419-PHY. 
Z.Q.Y. is supported by National Natural Science Foundation of China under Grant No. 61435007.

%

\end{document}